\documentclass[journal,10pt]{IEEEtran}

\usepackage[cmex10]{amsmath}
\usepackage[utf8]{inputenc}
\usepackage[pdftex]{graphicx}
\usepackage{soul,color}
\usepackage{epstopdf}
\usepackage{array}
\usepackage{floatrow}
\usepackage{caption}
\usepackage{subcaption}
\usepackage{url}
\usepackage{balance}
\usepackage{enumerate} 
\usepackage{amsbsy}
\usepackage[normalem]{ulem}
\usepackage{textcomp}
\usepackage{siunitx}
\usepackage{balance}
\usepackage{placeins}
\usepackage{tabularx, ragged2e}
\usepackage{listings}
\usepackage[dvipsnames]{xcolor}
\usepackage[T1]{fontenc}
\usepackage{helvet}
\usepackage{comment}

\def\BibTeX{{\rm B\kern-.05em{\sc i\kern-.025em b}\kern-.08em
    T\kern-.1667em\lower.7ex\hbox{E}\kern-.125emX}}
    
\usepackage{xspace}

\DeclareTextFontCommand{\textcomputer}{\fontfamily{cmr}\selectfont}

\begin{document}

\title{Demystifying VEINS: A Reality Check Against Living Lab Experiments}

\author{Antonio~Solida\IEEEauthorrefmark{1}, Giovanni~Gambigliani~Zoccoli\IEEEauthorrefmark{1}\IEEEauthorrefmark{2}, Gaetano~Orazio~Cauchi\IEEEauthorrefmark{1}, Filip~Valgimigli\IEEEauthorrefmark{1}, Salvatore~Iandolo\IEEEauthorrefmark{1}, Martin~Klapez\IEEEauthorrefmark{1}, Maurizio~Casoni\IEEEauthorrefmark{1}, Mirco~Marchetti\IEEEauthorrefmark{1} and Carlo~Augusto~Grazia\IEEEauthorrefmark{1}\IEEEauthorrefmark{3}\\[0.4cm]
\makebox[\textwidth][c]{%
  \begin{minipage}[t]{0.40\textwidth}
    \centering
    \IEEEauthorrefmark{1}\textit{Department of Engineering ``Enzo Ferrari''}\\
    \textit{University of Modena and Reggio Emilia}\\
    \{name.surname\}@unimore.it
  \end{minipage}%
  \begin{minipage}[t]{0.30\textwidth}
    \centering
    \IEEEauthorrefmark{2}\textit{InfoSystem Security S.R.L}\\
    ggz@infosystemsecurity.com
  \end{minipage}%
  \begin{minipage}[t]{0.30\textwidth}
    \centering
    \IEEEauthorrefmark{3}\textit{HIPERT S.R.L}\\
    carloaugusto.grazia@hipert.it
  \end{minipage}%
}
}

\maketitle

\begin{abstract}

Safety applications in vehicle-to-everything communications and Cooperative Intelligent Transport Systems rely on reliable and timely message exchange, which in turn depends on accurate modeling of wireless signal propagation. Simulation frameworks such as VEINS are widely adopted to design and evaluate such systems before deployment; however, their realism strongly depends on the validity of the underlying channel and antenna models. This work presents an empirical validation of the VEINS simulator against real-world data collected from the MASA living laboratory. Using the default configuration, we compare Received Signal Strength Indicator (RSSI), number of messages, and attenuation of the signal. The results show that VEINS systematically overestimates the RSSI value, while losing approximately $18\%$ of the total number of messages received compared to the MASA, revealing inconsistencies between simulation and reality. The contribution of this study is a direct comparison between simulated and real world data, establishing a quantitative basis for future calibration of VEINS parameters to improve the fidelity of VANET simulations in C-ITS safety research.

\end{abstract}


\makeatletter{\renewcommand*{\@makefnmark}{}
\footnotetext{This is a preprint version, the paper has been accepted for publication in the proceedings of IEEE Vehicular Technology Conference Spring 2026. The final published version will be available through IEEE.}\makeatother}

\section{Introduction}
\label{sec_intro}


In recent years, significant efforts have been devoted to enhancing road safety through the development of Cooperative Intelligent Transport Systems (C-ITS), which encompass a wide range of applications and services aimed at improving driver and passenger safety and traffic efficiency. The progress toward realizing C-ITS has mainly been driven by advances in Vehicular Ad Hoc Networks (VANETs), which support the rapid, direct exchange of information required by most cooperative applications; VANETs facilitate vehicle-to-everything (V2X) communication, allowing vehicles to exchange messages with other vehicles, roadside infrastructure, and network entities. These interactions rely on standardized wireless communication protocols that define key parameters such as coverage, data rate, latency, and security. Nevertheless, reliable data delivery remains a major challenge due to the highly dynamic topology of vehicular networks, frequent signal disruptions, and the transient nature of node connectivity.

 \begin{figure*}
    \centering
    \includegraphics[width=0.85\textwidth, trim = {0cm 0cm 0cm 0cm}]{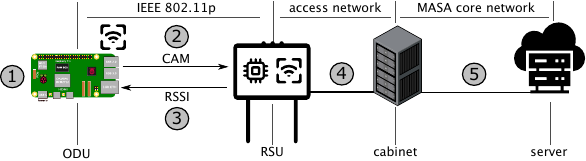}
    \caption{Setup scheme}
    \label{fig:rsu_to_odu}
 \end{figure*}

However, deploying and testing VANETs in real-world conditions is labor-intensive and costly, which is why most researchers rely on simulation tools for their evaluations. The precision of such studies depends heavily on the underlying mobility models, which must closely approximate real vehicular movement to yield meaningful results. Existing VANET simulators can generally be categorized into three main groups: mobility generators, such as SUMO, which model realistic traffic flow and vehicle behavior; network simulators, such as ns-2, ns-3, and OMNeT++, which focus on communication protocols and networking performance; and integrated frameworks, which combine both aspects to enable end-to-end evaluations. Among these integrated solutions, VEINS has become one of the most widely adopted frameworks. It tightly couples the SUMO road traffic simulator, capable of modeling microscopic vehicle behaviors such as speed, acceleration, and route dynamics, with the OMNeT++ network simulator, known for its modularity and strong community support \cite{haidari2019veins}. This co-simulation approach enables synchronized vehicle mobility and network communication, making VEINS particularly well-suited to evaluate vehicular network protocols~\cite{gambigliani2025radar, zoccoli2023vanets, zoccoli2022sixpack}; its popularity stems from its stability, portability, and realistic mobility representation, which have made it a de facto reference tool in VANET research.

Despite their widespread adoption and modular design, current VANET simulators, such as VEINS, exhibit significant discrepancies from real-world experimental results. Several studies have shown that even minor variations in physical-layer parameters, such as transmission power, fading, or antenna characteristics, can lead to substantial differences in simulated outcomes \cite{SPEIRAN2022359} \cite{takai2001effects}. These inconsistencies raise concerns about the extent to which the results obtained through such frameworks truly reflect the performance of real vehicular communication. In practice, simulators often rely on simplified propagation or interference models, idealized synchronization, and deterministic mobility traces, which fail to capture the stochastic nature of wireless channels and environmental factors observed in field experiments. Consequently, while VEINS remains a powerful and extensible research tool, its results must be interpreted with caution when used to derive conclusions about the reliability or safety of real-world systems. This gap between simulated and empirical performance motivates a critical re-evaluation of existing frameworks, emphasizing the need for calibration and hybrid validation approaches that combine simulation with controlled field testing.

In this context, the present work aims to bridge the gap between simulated and real-world vehicle network performance by comparing VEINS-based simulations with real tests conducted in the Modena Automotive Smart Area (MASA). This one provides a controlled urban environment with connected vehicles, smart roadside units, and 5G-enabled infrastructure, enabling reproducible and realistic testing of C-ITS technologies. The simulation campaign was designed to replicate the physical and environmental conditions of the MASA test site, including its road topology, traffic density, and communication configurations. By comparing simulation outcomes with measurements from real-world experiments, we quantify deviations between simulated and empirical results, identify the physical and protocol-level factors contributing to these discrepancies, and discuss calibration adjustments to enhance VEINS' fidelity. Ultimately, this study provides practical insights into the limitations of current VANET simulators and contributes to the development of more reliable, experimentally grounded simulation methodologies for future vehicular network research.

The paper is organized as follows.
Section~\ref{sec_related} provides an overview of related works.
Section~\ref{sec_method} presents the methodology used in this study, including the experimental setup and the test procedures, while Section~\ref{sec_testbed} describes the simulator setups.
Section~\ref{sec_test_results} discusses the test results and Section~\ref{sec_conclusions} concludes the article.


\section{Related Works}
\label{sec_related}

The work of Speiran et al. \cite{SPEIRAN2022359} analyzes the impact of various physical parameters, including transmission power, fading, propagation model, and noise level, on the performance of vehicular ad hoc networks (VANETs). The authors use simulations (e.g., in environments like VEINS) to evaluate metrics such as packet reception rate, latency, and coverage. The results show that different choices of physical models, for instance, between path loss and log-normal shadowing, can significantly affect simulation outcomes. The authors conclude that the strong sensitivity to physical parameters makes uncalibrated simulations potentially misleading, emphasizing the need for validation against real-world experimental data.

Altmemi et al. \cite{vanet_lento_2025} present a systematic review of the literature and a software-based evaluation of the VEINS framework in the context of VANET research. The authors analyzed studies published between 2011 and 2022 across three major scientific databases (IEEE Xplore, ScienceDirect and Scopus) and categorized them into three main areas: security, safety, and other VANET implementations. Through this classification, the study highlights key research gaps, including scalability challenges, computational overhead in large-scale simulations, and limited integration of emerging technologies such as 5G, blockchain and AI-driven optimization. Furthermore, the authors emphasize the lack of real-world validation of VEINS-based simulations, raising concerns about the applicability of the framework to real vehicular networking scenarios. Although VEINS remains a versatile tool, several limitations must be addressed to enhance its scalability, adaptability, and relevance, particularly in supporting next-generation paradigms such as edge computing, intelligent decision-making, and ultra-low-latency communications in 5G environments. However, the limitations identified in their analysis do not affect our work, as our study focuses on different performance aspects and does not rely on the large-scale or next-generation features they discuss.



Weber et al. \cite{9345525} identify several open challenges in current VANET simulators despite their extensive functionality. The authors highlight the need for improvements in several areas to enhance realism, reliability, and applicability. Current simulators such as VEINS and VENTOS only partially support security mechanisms and are not compliant with major standards (IEEE 1609.2 and ETSI ITS), which prevents comparing novel security solutions with existing specifications; extending simulators to support these standards fully requires significant research in systems and programming. Furthermore, no general-purpose tools exist for fault injection in VANETs, which limits the ability to assess system dependability under hardware, software, or data faults and to establish safety benchmarks. Coupling VANET simulators with real-time hardware in Hardware-in-the-Loop setups remains challenging due to performance constraints, particularly when simulating cryptographic operations, necessitating further research for efficient integration without compromising fidelity. Although simulators now include more realistic propagation and environmental models, emerging technologies such as 5G, edge computing, and UAVs introduce conditions that are not yet accurately represented, indicating the need to extend simulators to cover these effects. Finally, the increasing complexity of simulators makes debugging and validation more difficult, highlighting the importance of developing context-driven automated testing frameworks to facilitate maintenance, accelerate development, and improve software reliability. In general, the authors emphasize that future VANET simulators must evolve to ensure compliance with standards, enable realistic, fault-tolerant modeling, and support efficient, real-time, automated testing.

Recent studies addressing misbehavior detection in V2X highlight the importance of physical-level validations. For example, So et al. \cite{so2019physical} propose three plausibility checks based on the Received Signal Strength Indicator (RSSI) to detect location spoofing and report a detection rate of up to 83.7\% with 95.9\% accuracy on the VeReMi dataset. These approaches exploit correlations between received power and declared distance and work independently of the correctness of the majority of nodes. This evidence underscores that effective countermeasures require accurate physical channel and signal power models, not just application-level checks. In this context, our analysis of VEINS shows that current physical-layer modeling (RSSI, fading, antenna pattern, environmental attenuation) is often insufficient to evaluate algorithms that exploit real signal behavior correctly; consequently, promising results obtained in VeReMi or other datasets may not be replicated in VEINS without appropriate extensions and calibrations.

Most existing studies on Misbehavior Detection (MBD) in Cooperative Intelligent Transport Systems (C-ITS) rely on simulation-based evaluation rather than real-world experimentation. As noted by Sakiz and Sen \cite{sakiz2017survey}, the majority of MBD mechanisms proposed for VANETs have been validated only in simulation environments, making the reliability of vehicular simulators a key factor in assessing their credibility. Traditional tools such as Veins and VENTOS are widely adopted for communication and mobility analysis but provide limited capabilities for modeling misbehavior and evaluating security mechanisms under realistic adversarial conditions. To address these limitations, Kamel et al. \cite{kamel2020simulation} introduced F2MD, an open-source framework designed explicitly for MBD research. F2MD integrates diverse attack models, detection techniques, pseudonym change strategies, and real-time visualization tools, offering reproducible benchmarks and extensible components for security-oriented experimentation. These efforts reflect a growing shift in the community toward behavior- and security-aware simulation environments that complement the communication-centric focus of traditional VANET simulators.





\section{Method}
\label{sec_method}
The Modena Automotive Smart Area (MASA) is an open-air urban laboratory for testing and validating intelligent and cooperative mobility technologies. Covering part of Modena’s urban fabric and university campus, MASA is equipped with digital infrastructure and sensors for real-time data collection, enabling realistic experimentation with connected and autonomous vehicles, V2X systems, and smart-road applications. ETSI-compliant Roadside Units (RSUs) are installed and connected to the area’s fiber-optic backbone to ensure minimal latency, implementing C-ITS functionalities based on IEEE 802.11p, 4G, and 5G for both vehicle-to-vehicle (V2V) and vehicle-to-infrastructure (V2I) communications.

Movyon Electronics’ RSUs comply with ITS-G5 (IEEE 802.11p) and 3GPP C-V2X (Releases 14 and 15) standards. Each unit integrates a dedicated ITS-G5 stack, operates in the 5.9 GHz band with reception sensitivity up to $-97$ dBm, and includes a multiconstellation GNSS receiver (GPS, Galileo, GLONASS, Beidou) for precise localization, including altitude. PCIe expansion allows scalable LTE/5G modules. A limitation is the lack of an interface for retrieving RSSI of transmitted packets; thus, RSSI was measured via the On-Board DSRC Unit (ODU) supporting IEEE 802.11p \cite{iandolo2025odu}.

The experimental evaluation involved a joint test of TX and RX performance between RSUs and ODUs. Downlink (RSU-to-ODU) measured RSSI, while uplink (ODU-to-RSU) assessed transmitted data and packet error rate (PER). Messages were ETSI-compliant CAMs with valid GPS data for georeferencing.

During the test (setup in Figure \ref{fig:rsu_to_odu}, path in Figure \ref{fig:sumo_distance_masa}), the ODU, mounted on a passenger vehicle with an external GNSS receiver, received CAM broadcasts at 100-ms intervals. Message handling used RSU APIs for precise V2X control. The ODU employed two high-performance omnidirectional antennas providing 360° coverage, reducing multipath and occlusions compared to standard lower-gain antennas.

All CAMs received by the ODU were stored in PCAP files via \texttt{tcpdump} on a monitor-mode interface, allowing extraction of per-packet RSSI and GPS coordinates for accurate post-processing and spatial correlation.


 
\section{Software Setup}
\label{sec_testbed}


To compare the data extracted from the real-world scenario of the MASA described in Section~\ref{sec_method}, the same scenario was replicated in a simulator environment using the SUMO (Simulation of Urban Mobility)~\cite{SUMO2018}, OMNeT++~\cite{varga2008overview}, and VEINS~\cite{Veins} framework. This hybrid platform enables the joint analysis of vehicular mobility and communications in VANET scenarios. SUMO is an open-source microscopic traffic simulator that reproduces individual vehicle behavior, including dynamics such as acceleration, deceleration, interactions with other vehicles, traffic light management, and compliance with traffic rules. It allows real road networks to be imported from cartographic sources (such as OpenStreetMap), and complex, realistic traffic patterns to be generated. OMNeT++ is a discrete event simulator widely used in academia for modeling communication networks, thanks to its modular architecture and the ability to define custom protocols, network stacks, and wireless scenarios. VEINS acts as middleware between SUMO and OMNeT++, implementing a bidirectional coupling through the TraCI (Traffic Control Interface) interface. This allows for real-time exchange of information between the two simulators: traffic conditions influence network parameters (e.g., connectivity or node density), while communication events—such as the reception of cooperative or safety messages—can modify vehicle behavior, including speed, trajectory, and routing decisions.

By default, VEINS uses the Free Space Path Loss (FSPL) model to compute signal attenuation as a function of distance and frequency, assuming unobstructed line-of-sight propagation~\cite{VeinsObstacle2}. Each vehicle is equipped with an isotropic antenna that provides uniform radiation in all directions, with a carrier frequency of 5.9 GHz, corresponding to IEEE 802.11p vehicular communications. Packet reception depends on the Signal-to-Interference-plus-Noise Ratio (SINR) remaining above a defined threshold throughout the frame duration. As vehicles move in SUMO, OMNeT++ updates antenna positions and orientations, and the Received Signal Strength Indicator (RSSI) is computed dynamically, decreasing with distance in accordance with the theoretical free-space propagation model.

To replicate the real-world scenario inside the simulator, the MASA living laboratory area was imported using the integrated tools available in SUMO. The 17 RSUs in the MASA area were placed by converting their GPS coordinates to the SUMO reference system. These RSUs were configured to transmit messages every 100 ms, replicating the configuration of the real experiment. The simulations use the default VEINS parameters~\cite{VEREMI,LUST}, with key values configured as follows:
\begin{itemize}
    \item \textit{Transmission power}: $20~mW$ (approximately $13~dBm$), representing the sending power of all network cards in the simulation.
    \item \textit{Lowest power level}: $-89~dBm$, used by the decider layer of the receiver entities to determine whether the messages have enough power to pass through the next layer of the stack.
    \item \textit{Noise floor}: $-98~dBm$, representing ambient interference that is used to realistically model the propagation behavior of radio signals and contribute to the calculation of the signal-to-noise ratio at the receiver.
    \item \textit{Building attenuation}: $9~dBm$, used to compute the attenuation of buildings while the signal travels through them.
    \item \textit{Distance attenuation}: $0.4~dBm$, used to calculate the interference caused by the distance traveled by the messages to reach the receiver antenna, modeling the propagation of information in a vehicular network.
\end{itemize}

To precisely reproduce the vehicle’s route, the GPS positions collected during the MASA test were used to create a SUMO trace using the \texttt{TraceMapper} utility of SUMO. The \texttt{TraceMapper} utility can map a sequence of geographic coordinates to a contiguous sequence of network edges, producing a valid route within the given network.

To avoid replicating the real traffic condition, a single vehicle was simulated, and all traffic lights were removed from the SUMO scenario, so that the vehicle was not constrained by SUMO’s built-in car-following and signal-compliance logic.
However, a marked divergence, in terms of speed and time, was observed between the simulated vehicle and the real one, since the simulation was carried out without taking into account the slowdowns caused by the other vehicles observed during the MASA test.
To recreate the same path considering speed and time, the real speed value collected during the MASA test was enforced at regular time interval using the distance–time equation $s = \frac{d}{t}$.
Where $d$ represents the distance between two consecutive coordinates $(C_x, C_y)$ and $(N_x, N_y)$, while $t$ is obtained as the difference between the two corresponding timestamps (approximately $100~ms$).

\begin{figure}[htb]
  \centering
    \includegraphics[width=0.98\textwidth]{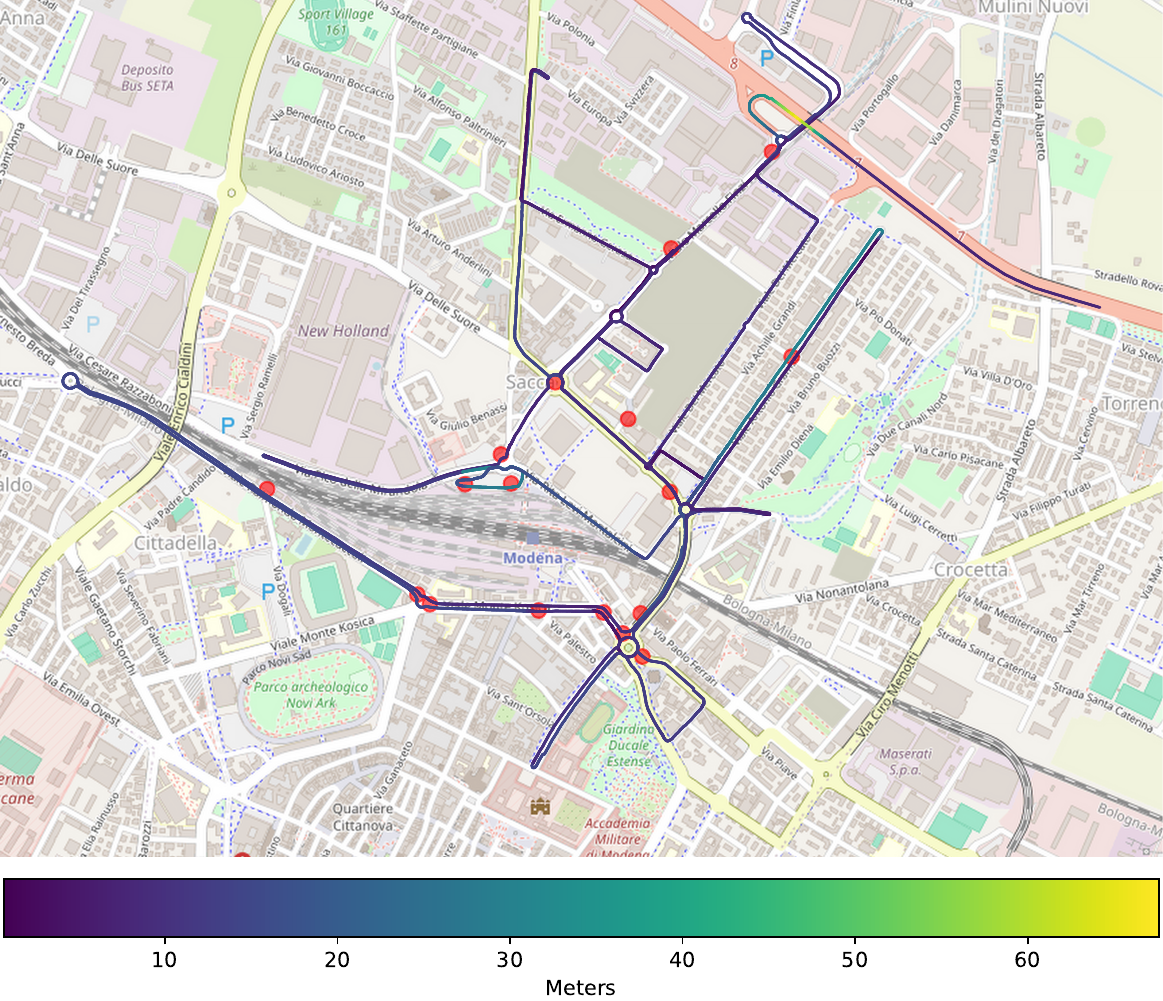}
    \caption{RSUs displacement and distance between \emph{VEINS} vehicle and \emph{MASA} vehicle}
    \label{fig:sumo_distance_masa}
\end{figure}

The results of these adjustments are shown in Figure~\ref{fig:sumo_distance_masa}, where the RSU positions are depicted in red, while the vehicle’s path is colored according to its distance from the real vehicle positions. As can be observed, a maximum distance of approximately $50~m$ is reached. However, along the route, the distances generally fluctuate between $10$ and $20~meters$, meaning that the vehicle's trajectory was accurately replicated considering the distance between the real vehicle and the simulated one.

During the simulation, the vehicle was configured to collect all messages originated from the RSUs. For each message received, the following data were stored: the RSU identifier (station ID), the RSU position, the GPS position of the vehicle, and the RSSI value.

\section{Test Results}
\label{sec_test_results}

This Section presents the results obtained from the analysis of the data extracted from the simulation setup described in Section~\ref{sec_testbed}. An initial analysis of the results consist of evaluating the RSSI distributions among the real and the simulated scenario.

\begin{figure*}
  \centering
    \includegraphics[width=0.99\textwidth, trim = {0cm 0.2cm 0cm 0cm}]{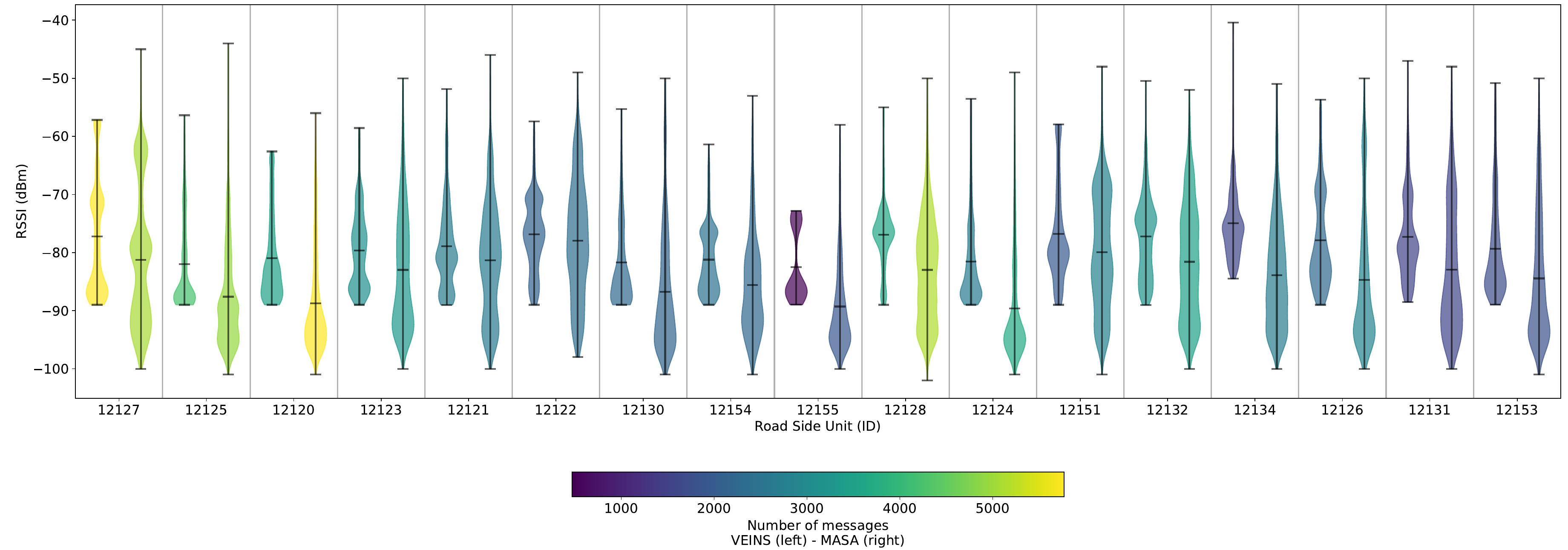}
    \caption{Comparison between number of messages and RSSI value between MASA and VEINS}
    \label{fig:RSSI_violin}
\end{figure*}

Figure~\ref{fig:RSSI_violin} shows a set of violin plots, arranged horizontally by RSU. Along the $x$-axis the RSUs are labeled using their station IDs while along the $y$-axis the RSSI value is reported using $dBm$, with fewer negative values indicating a stronger signal. For each station ID, two adjacent violins are shown to enable comparison between simulated and real measurements: the violin on the \emph{left} corresponds to data produced in \emph{VEINS}, while the violin on the \emph{right} corresponds to measurements collected in the \emph{MASA} laboratory. 
Each violin visualizes the full distribution of RSSI values observed for that RSU, appearing wider where observations are denser and narrower where observations are sparse; thin tails indicate low-probability extremes. The color is used to encode the number of messages received, starting from a dark purple indicating about $1000$ messages reaching yellow which represent more than $5000$ number of messages received.

Figure~\ref{fig:RSSI_violin} is intended to be read group by group along the $x$-axis. Within each group, the left (\emph{VEINS}) and right (\emph{MASA}) violins are compared to assess how closely the simulation reproduces real-world conditions.
Focusing on the $x$-axis, a consistent pattern can be observed: the \emph{MASA} distributions generally exhibit lower (more negative) RSSI values compared to the \emph{VEINS} distributions, while the number of received messages, indicated by the color, is more similar across the two scenarios, with notable exceptions in IDs $12127$, $12120$, $12155$ and $12128$, where appreciable differences can be noted. While opposite trend can be observed considering the RSU $12127$ and $12120$, respectively.
However, when considering the entire vehicle route, a difference of approximately $18\%$ in favor of \emph{MASA} was observed, indicating that the real-world system achieves a higher overall message delivery rate compared to the \emph{VEINS} simulation environment.

\begin{figure*}[ht]
    \centering
    \begin{subfigure}[t]{.48\textwidth}
        \centering
        \includegraphics[width=0.98\linewidth]{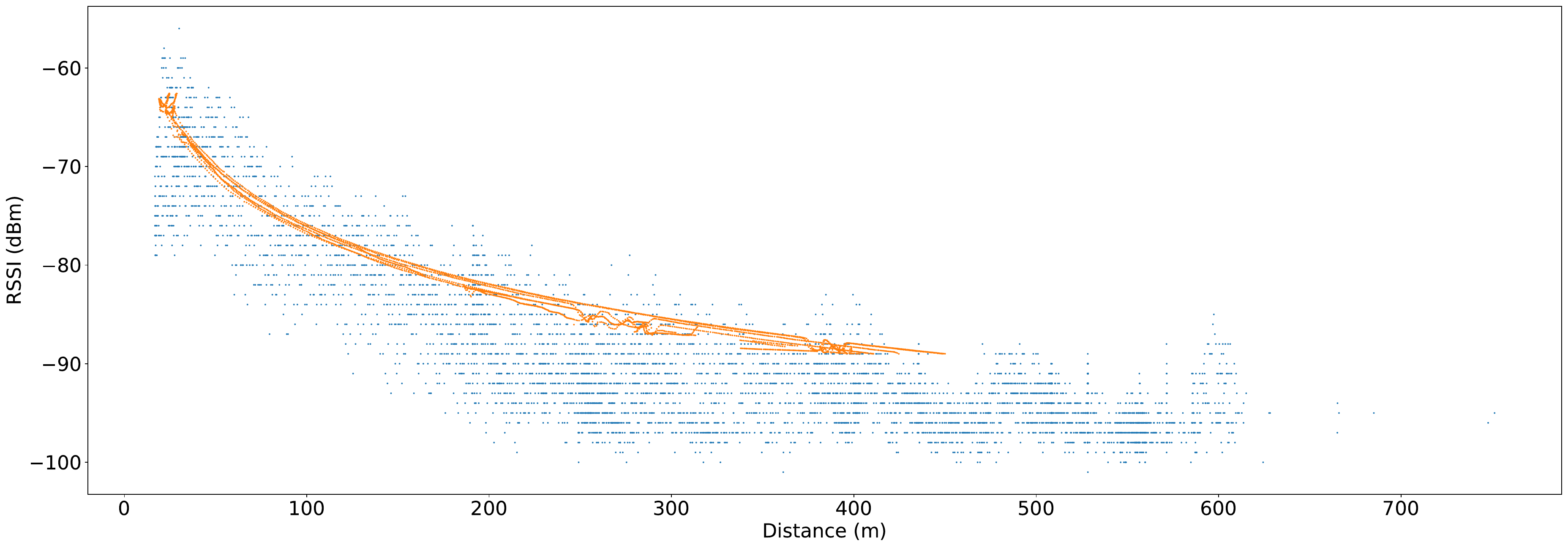}
        \caption{RSU ID~\texttt{12120}}
        \label{fig:RSSI_12120}
    \end{subfigure}
    \hfill
    \begin{subfigure}[t]{.48\textwidth}
        \centering
        \includegraphics[width=0.98\textwidth]{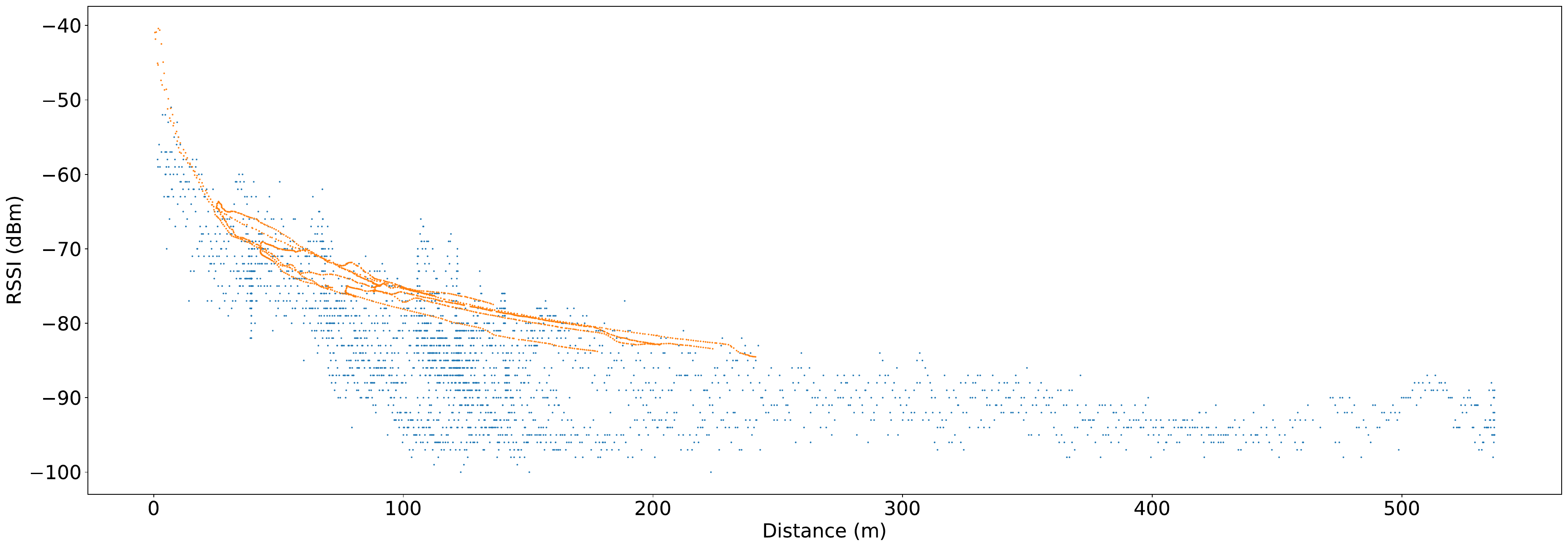}
        \caption{RSU ID~\texttt{12134}}
        \label{fig:RSSI_12134}
    \end{subfigure}
    \hfill
    \begin{subfigure}[t]{.48\textwidth}
        \centering
        \includegraphics[width=0.98\textwidth, trim = {0cm 0.5cm 0cm 0cm}]{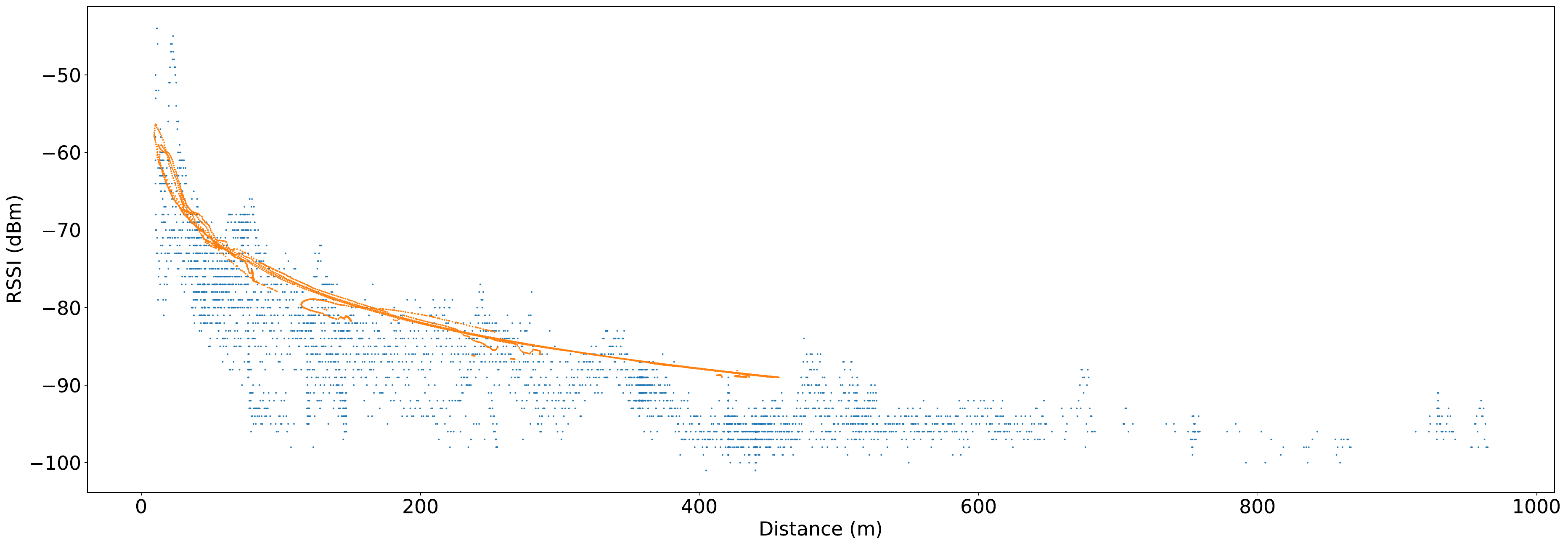}
        \caption{RSU ID~\texttt{12125}}
        \label{fig:RSSI_12125}
    \end{subfigure}
    \hfill
    \begin{subfigure}[t]{.48\textwidth}
        \centering
        \includegraphics[width=0.98\textwidth, trim = {0cm 0.5cm 0cm 0cm}]{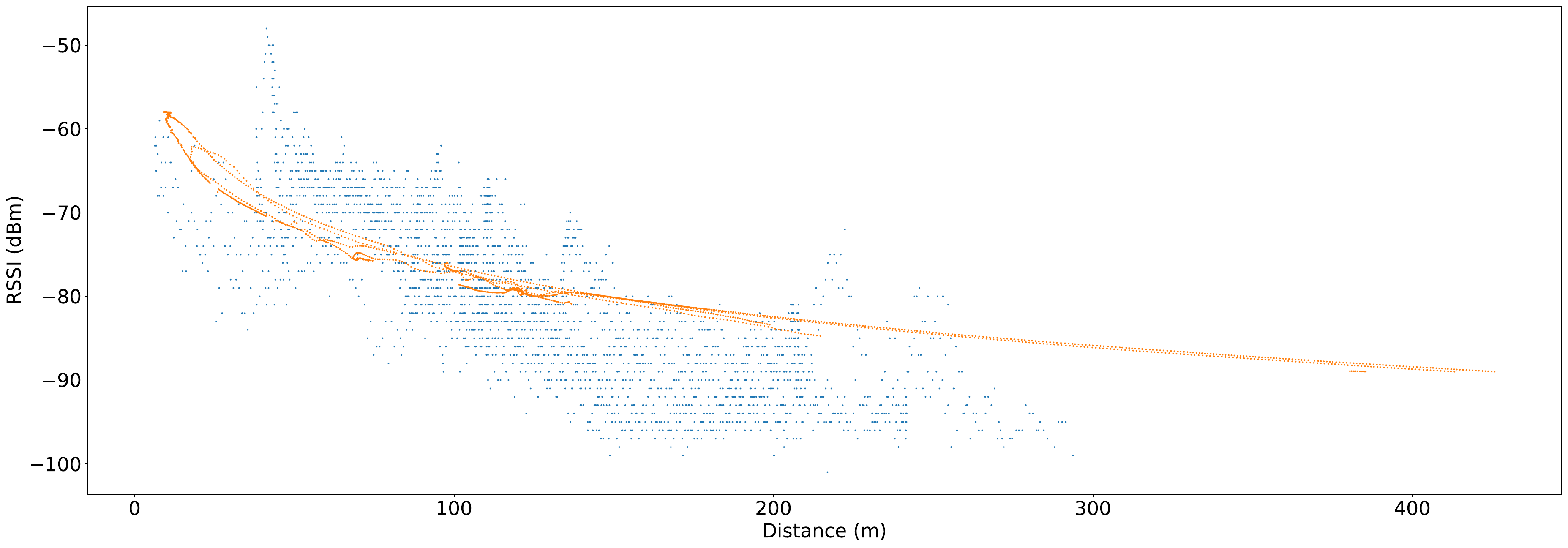}
        \caption{RSU ID~\texttt{12151}}
        \label{fig:RSSI_12151}
    \end{subfigure}

    \caption{RSSI value over distance}
    \label{fig:RSSI_over_distance}
\end{figure*}

Figure~\ref{fig:RSSI_over_distance} presents the RSSI as a function of distance for the $4$ representative RSUs. The distance is reported in meters on the $x$-axis, while on the $y$-axis the RSSI is reported in $dBm$. For each figure, two point clouds are shown in each figure: the blue markers correspond to measurements collected in the \emph{MASA}, while the orange markers correspond to the \emph{VEINS} simulation.

From a general point of view, a decreasing trend of RSSI with increasing distance is visible in all cases, as expected from radio propagation. However, the \emph{VEINS} pattern follows a smooth, nearly monotonic envelope with limited scatter, reflecting a simplified propagation model. In contrast, the \emph{MASA} measurements are markedly more variable: the point clouds are wider at comparable distances, local fluctuations and clusters appear, and gaps indicate intervals with missed receptions or transient attenuation.

Two recurring patterns emerge across the selected RSUs. First, at similar distances, the MASA data generally denote lower RSSI levels than in the simulation, indicating real-world losses that the \emph{VEINS} simulator does not fully capture (as also shown in Figure~\ref{fig:RSSI_violin}). Second, a non-negligible difference can be observed in the number of received messages, revealing an imbalance that further highlights the discrepancies between the simulation and reality.
Taken together, the figures suggest that \emph{VEINS} accurately reproduces the expected distance-dependent decay, but tends to smooth out real-world variability.

In almost all the images in Figure~\ref{fig:RSSI_over_distance} (three over four), we can observe that the \emph{MASA} test collects more messages compared to \emph{VEINS}, confirming the difference of about $18\%$ in terms of the number of messages received.
In contrast, Figure~\ref{fig:RSSI_12151} shows a different trend, the RSU with station ID equal to \texttt{12151} is able to collect more messages inside the \emph{VEINS} simulator with respect to the \emph{MASA}, meaning that the interference and attenuation of the \emph{VEINS} simulator struggle to replicate the real world scenario of the MASA laboratory.

\section{Conclusions}
\label{sec_conclusions}
This work presented a comparison between the VEINS simulation framework (composed of SUMO and OMNeT++) and a real-world deployment. The objective was to assess how the simulator is able to reproduce real vehicular communication conditions.

The analysis shows that the simulator can reproduce the general trends observed in reality: the message exchange dynamics and the general behavior of the network are reasonably captured. This confirms that VEINS remains a useful tool for the preliminary evaluation of vehicular network solutions, where repeatability and control of parameters are required.

However, the results highlight a non-negligible gap between simulation and real measurements. In particular, VEINS reports a $-17.88\%$ difference in the total number of messages received when considering all RSUs, compared to the real data collected in the MASA area. Moreover, in terms of the received signal power, the simulator yields a mean value of $-78.96~dBm$, while the real MASA campaign shows a lower mean power of $-84.44~dBm$ highlighting a more simplified propagation models, idealized environmental conditions, or less interference in the VEINS simulator.

As a consequence, while simulation is effective for relative comparisons and analyzes, it should not be considered a perfect substitute for field trials.

Future work will focus on reducing this gap by (i) refining the propagation model to better reflect the measurement environment and (ii) tuning VEINS parameters using real data as ground truth.
This calibration step is essential to improve the fidelity of VEINS when it is used to predict performance in realistic deployment scenarios.

\section*{Acknowledgements}
This work is part of the FRODDO project and has been co-funded by the European Union's Horizon Europe Research and Innovation Programme under Grant Agreement No. 101147819.

\balance
\bibliographystyle{IEEEtran}
\bibliography{wm24}

\end{document}